\documentclass[a4paper]{jpconf}
\usepackage{subfigure}
\usepackage{amssymb}
\usepackage{axodraw}
\usepackage{graphicx}

\begin{document}
\vspace*{-2cm}
\noindent
\hspace*{13.5cm}IPPP/07/03\\
\hspace*{13.5cm}DCPT/07/06

\vspace*{0cm}
\title{Axions, their Relatives and Prospects for the Future}

\author{Joerg Jaeckel}

\address{Centre for Particle Theory, University of Durham, Durham, DH1 3LE, UK}

\ead{joerg.jaeckel@durham.ac.uk}

\begin{abstract}
The observation of a non-vanishing rotation of linear polarized laser light after passage through a strong magnetic field by the PVLAS collaboration has renewed the interest in light particles coupled to photons.
Axions are a species of such particles that is theoretically well motivated.
However, the relation between coupling and mass predicted by standard axion models conflicts with the PVLAS observation.
Moreover, light particles with a coupling to photons of the strength required to explain PVLAS face
trouble from astrophysical bounds.
We discuss models that can avoid these bounds.
Finally, we present some ideas to test these possible explanations of PVLAS experimentally.
\end{abstract}

\section{Motivation -- the PVLAS Observation}
Recently, the PVLAS collaboration reported the observation of a non-vanishing rotation of linear polarized laser light after passage through a strong transverse magnetic field \cite{Zavattini:2005tm},
\begin{equation}
 \Delta\theta\sim 10^{-12}\, \rm{rad}
\end{equation}
per pass through a magnetic field of $B\approx 5\,\rm{T}$ and length $L=1\,\rm{m}$ and using a laser with
wavelength $\lambda\sim 1000\,\rm{nm}$.
Such a signal is not expected in standard QED with electrons. If confirmed this could be the first direct evidence
for physics beyond the standard model.

\begin{figure}
\vspace{1cm}
\hspace{-1.5cm}
\subfigure[]{
\scalebox{1.2}[1.2]{
\begin{picture}(100,100)(0,0)
\includegraphics[width=9cm]{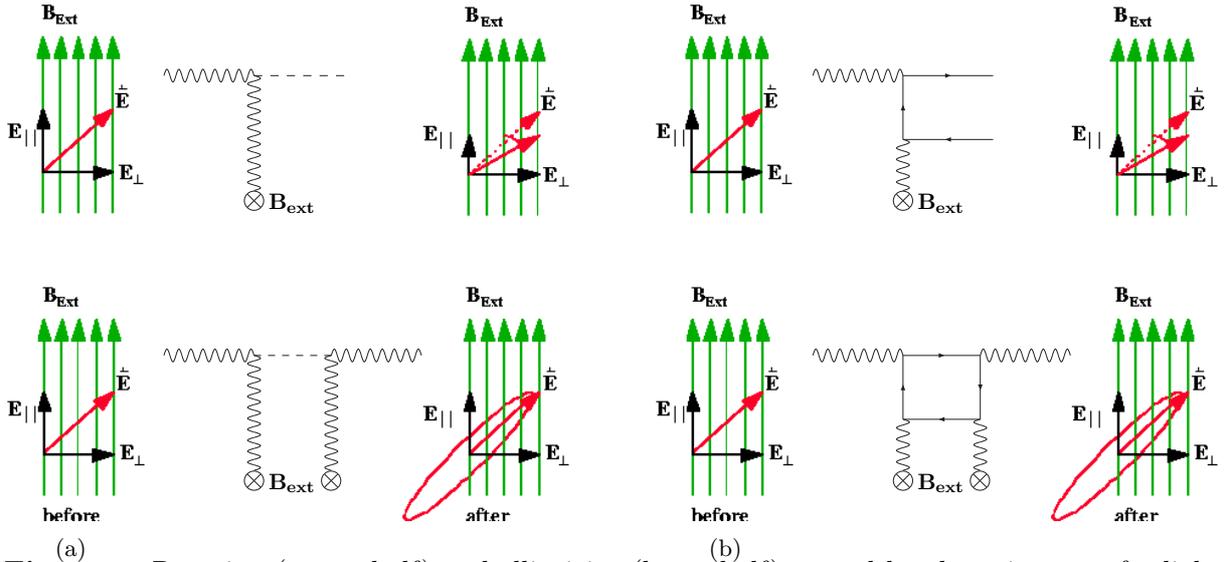}
\scalebox{0.4}[0.4]{
\SetOffset(-460,260)
\Text(100,-10)[c]{\scalebox{1.6}[1.6]{$\mathbf{B}_{\mathbf{ext}}$}}
\Photon(0,90)(70,90){5}{7.5}
\Photon(70,90)(70,00){5}{10.5}
\DashLine(70,90)(140,90){6}
\SetOffset(-460,262)
\CArc(70,-10)(7.5,0,360)
\Line(75,-5)(65,-15)
\Line(65,-5)(75,-15)
\SetOffset(-460,40)
\Text(100,-10)[c]{\scalebox{1.6}[1.6]{$\mathbf{B}_{\mathbf{ext}}$}}
\Photon(0,90)(70,90){5}{7.5}
\Photon(70,90)(70,0){5}{10.5}
\DashLine(70,90)(130,90){6}
\SetOffset(-460,42)
\CArc(70,-10)(7.5,0,360)
\Line(75,-5)(65,-15)
\Line(65,-5)(75,-15)
\SetOffset(-460,40)
\Photon(130,90)(130,0){-5}{10.5}
\SetOffset(-400,42)
\CArc(70,-10)(7.5,0,360)
\Line(75,-5)(65,-15)
\Line(65,-5)(75,-15)
\SetOffset(-460,40)
\Photon(130,90)(200,90){5}{7.5}}
\end{picture}}
\label{alpconversion}}
\hspace{3.8cm}
\subfigure[]{
\scalebox{1.2}[1.2]{\begin{picture}(100,100)(0,0)
\includegraphics[width=9cm]{stream.eps}
\scalebox{0.4}[0.4]{
\SetOffset(-460,260)
\Text(100,-10)[c]{\scalebox{1.6}[1.6]{$\mathbf{B}_{\mathbf{ext}}$}}
\Photon(0,90)(70,90){5}{7.5}
\Photon(70,40)(70,0){5}{4.5}
\ArrowLine(70,90)(140,90)
\ArrowLine(70,40)(70,90)
\ArrowLine(140,40)(70,40)
\SetOffset(-460,262)
\CArc(70,-10)(7.5,0,360)
\Line(75,-5)(65,-15)
\Line(65,-5)(75,-15)
\SetOffset(-460,40)
\Text(100,-10)[c]{\scalebox{1.6}[1.6]{$\mathbf{B}_{\mathbf{ext}}$}}
\Photon(0,90)(70,90){5}{7.5}
\Photon(70,40)(70,0){5}{4.5}
\ArrowLine(70,90)(130,90)
\ArrowLine(70,40)(70,90)
\ArrowLine(130,40)(70,40)
\ArrowLine(130,90)(130,40)
\SetOffset(-460,42)
\CArc(70,-10)(7.5,0,360)
\Line(75,-5)(65,-15)
\Line(65,-5)(75,-15)
\SetOffset(-460,40)
\Photon(130,40)(130,0){-5}{4.5}
\SetOffset(-400,42)
\CArc(70,-10)(7.5,0,360)
\Line(75,-5)(65,-15)
\Line(65,-5)(75,-15)
\SetOffset(-460,40)
\Photon(130,90)(200,90){5}{7.5}}
\end{picture}}
\label{milliconversion}}
\vspace{-0.6cm}
\caption{Rotation (upper half) and ellipticity (lower half) caused by the existence of a light neutral spin-0 boson (left) or a light particle with a small electric charge (right) (figure adapted from \cite{Brandi:2000ty}).}
\label{fig:ph_reg}
\end{figure}

One obvious possible explanation, and indeed the one which was also a
motivation for the BFRT \cite{Semertzidis:1990qc,Cameron:1993mr} and PVLAS experiments, may be offered by the
existence of a new light neutral spin-$0$ boson
$\phi$~\cite{Maiani:1986md} that could be a scalar or a pseudoscalar.
If this particle is, e.g., a pseudoscalar the interaction reads
\begin{equation}
\label{pseudoscalar}
{\mathcal L}_{\rm{int}}
  =-\frac{1}{4}g\phi F_{\mu\nu}\widetilde{F}^{\mu\nu}
  =g\phi (\vec{E}\cdot\vec{B}).
\end{equation}
In a homogeneous magnetic background $\vec{B}$, this interaction can convert the laser photons into pseudoscalars (upper part of Fig. \ref{alpconversion}).
The leading order contribution to this process comes from the
term $\sim \vec{E}_{\gamma}\cdot\vec{B}$.
The polarization of a photon is given by the direction of the
electric field of the photon, $\vec{E}_\gamma$.
Therefore, only those fields polarized
parallel to the background magnetic field will have nonvanishing
$\vec{E}_{\gamma}\cdot\vec{B}\neq0$ and interact with the
\mbox{pseudoscalar} particles. As depicted in Fig. \ref{alpconversion} this leads to an
absorption of photons with a polarization parallel to the magnetic field and this in turn leads
to an overall rotation of the polarization.

The rotation observed by PVLAS can be reconciled with the non-observation
of a rotation and ellipticity by the earlier BFRT experiment \cite{Semertzidis:1990qc,Cameron:1993mr}, if the hypothetical neutral
boson has a mass and a coupling
to two photons in the range
\begin{equation}
\label{masscoupling}
m_\phi\sim (1-1.5)\,\rm{meV}
\quad\quad
g\sim (1.7-5.0)\times 10^{-6}\,\rm{GeV}^{-1}.
\end{equation}

In addition to the rotation preliminary data \cite{PVLASICHEP,Cantatore:IDM2006}
presented at various conferences suggests the existence of a small ellipticity,
\begin{equation}
 \psi\sim 10^{-12} \,\rm{rad}
\end{equation}
per pass through the magnetic field (parameters as above). Such a signal could be explained by the virtual production of particles as depicted in the lower part of Fig. \ref{alpconversion}. Again, only the photons polarized parallel to the magnetic field are affected. Roughly speaking those are a bit delayed because the massive intermediate particle is a bit slower. This results in a phase difference between the polarizations parallel and perpendicular to the magnetic field which appears as an ellipticity in the final polarization.
This allows an independent calculation of the required mass $m_{\phi}$ using only PVLAS data. The results are consistent with Eq. (\ref{masscoupling}) \cite{Ahlers:2006iz}.

An alternative \cite{Gies:2006ca} is the existence of light particles with a small electric charge. These are often called millicharged particles although their charge may be significantly less than $10^{-3}$.
They could be bosons or fermions. For example, for fermions the interaction part of the Lagrangian would be
\begin{equation}
\label{lintDsp}
{\mathcal L}_{\rm int}^{\rm Dsp}= \epsilon\, e\,
\overline{\psi}_{\epsilon}
\gamma_\mu \psi_{\epsilon} A^\mu,
\end{equation}
with $\psi_\epsilon$ being a Dirac spinor {(``Dsp'')}.
As depicted in Fig. \ref{milliconversion}
the explanation of a possible rotation and ellipticity signal works in complete analogy to the case of spin-0
particles only pairproduction now replaces single particle production. Although it is not immediately obvious from the interaction Lagrangian, the rate for pairproduction and the birefringence caused by virtual pairproduction
do again depend on the relative orientation of laser polarization and background magnetic
field \cite{Toll:1952,Klepikov:1954,Erber:1966vv,Baier:1967,Klein:1968,Adler:1971wn,Tsai:1974fa,Daugherty:1984tr}.

It is clear that millicharged particles, too, must be light to allow for the rotation observed in PVLAS.
Real production of pairs requires that the total mass of the produced particles is less than the energy
of the laser photon $2m_{\epsilon}\leq \omega_{\gamma}$ since the (constant) background field cannot
contribute to the energy. Furthermore the charge must be sufficiently small to avoid conflicts with other
existing laboratory
experiments\footnote{We willl discuss astrophysical bounds later in Sect. \ref{astro}.}
$\epsilon\lesssim10^{-5}$ \cite{Mitsui:1993ha,Badertscher:2006fm,Prinz:1998ua,Gninenko:2006fi}.
Consistency with PVLAS, BFRT and the recent Q\&A measurements \cite{Chen:2006cd}
requires \cite{Ahlers:2006iz,Gies:2006ca},
\begin{equation}
 m_{\epsilon}\lesssim 0.1\, \rm{eV} \quad\quad \epsilon\sim 10^{-6}.
\end{equation}

\section{Axions in a Nutshell}
One light particle species coupled to photons is particularly well motivated by
theory: axions \cite{Weinberg:1977ma,Wilczek:1977pj,Peccei:1977hh,Peccei:1977ur}. Therefore, let us briefly review axions and discuss their (in)viability to explain PVLAS (for a more complete
review of axions see, e.g., \cite{Kim:1986ax,Svrcek:2006yi}).

Axions are motivated by the so called strong CP problem.
As the name indicates it has its origin in the theory of strong interactions, QCD.
Therefore, let us start by writing down the Lagrangian of QCD with quarks,
\begin{equation}
 S=\int d^{4}x \bigg[ -\frac{1}{4 g^{2}}G^{a,\mu\nu}G^{a}_{\mu\nu}-\frac{\theta}{32\pi^2}G^{a,\mu\nu}\tilde{G}^{a}_{\mu\nu}
 +\i \bar{\psi}D_{\mu}\gamma^{\mu}\psi+ \bar{\psi}{\rm{M}}\psi\bigg],
\end{equation}
where $G$ is the gluon field strength and $\tilde{G}$ its dual.
This Lagrangian contains a term $\sim \theta G \tilde{G}$ that is often absent in discussions of QCD. This term is perfectly allowed by gauge symmetry. So why is it often neglected? The reason is that this term violates CP in a strong manner but this is not observed in nature. More precisely,
\begin{equation}
\label{observ}
\theta+\rm{Arg} \det(M) \quad \rm{is~observable}
\end{equation}
and would appear in CP violating properties of hadrons as, e.g., an electric dipole moment of the neutron.
Measurements of the electric dipole moment of the neutron  \cite{Ramsay,Baker:2006ts} now constrain
\begin{equation}
 |\theta+\rm{Arg} \det(M)=0|<3\times10^{-10}.
\end{equation}

As mentioned in the beginning of this section $\theta$ is an allowed parameter which is undetermined in QCD.
Naturally it should be of order one. So the question is: why is it so small? This naturalness problem is the strong CP problem.

One prominent mechanism to solve this problem is the axion solution. The idea is to make $\theta$ a dynamical degree
of freedom with the following properties:
\begin{itemize}
 \item{} $\theta$ has no tree-level potential
 \item{} $\theta$ has only derivative couplings.
\end{itemize}
Choosing a basis such that $\rm{Arg} \det(M)=0$
one can then show (see, e.g., \cite{Kim:1986ax}) that for the effective potential,
\begin{equation}
\label{potential}
V[0]<V[\theta],
\end{equation}
i.e. the global minimum of the potential lies at the CP conserving value $\theta=0$, and $\theta$ will automatically evolve towards this value.

However, there is still the question from where this dynamical degree of freedom originates. Using
\begin{equation}
 \int d^{4}x \frac{G^{a,\mu\nu}\tilde{G}^{a}_{\mu\nu}}{32\pi^2}=n \in {\mathbb{Z}}
\end{equation}
one finds that $\theta$ is an angular degree of freedom
\begin{itemize}
 \item {} $\theta\in [0,2\pi)$.
\end{itemize}
Together with the two above properties this is reminiscent of a Goldstone boson of a spontaneously broken
U(1) symmetry.
Choosing a suitable axial U(1) symmetry one finds indeed that for the Goldstone
direction a coupling to $G\tilde{G}$ is generated via a triangle diagram. This is the famous
Peccei-Quinn symmetry \cite{Peccei:1977hh,Peccei:1977ur} and the particle is the
axion \cite{Weinberg:1977ma,Wilczek:1977pj}.

So far we have been talking about the axion as a Goldstone boson. However, Eq. (\ref{potential}) already shows that
the potential for the axion is not completely flat and the particle will have an effective mass.
Indeed, the Peccei-Quinn symmetry is anomalous and the axion is only a pseudo-Goldstone boson.

\enlargethispage{0.2cm}
A simple estimate (for more see, e.g., \cite{Weinberg:1977ma,Dine:1982,Sikivie:1982,Weinberg})
for this mass can be obtained by looking at the transformations
of the up and down quark condensates\footnote{For simplicity we consider only two flavors.}
under chiral rotations $\alpha$ and Peccei-Quinn
rotations\footnote{Effectively we use the Peccei-Quinn rotations to transfer the $\theta G\tilde{G}$-term
into complex quark masses: The rotations change $\rm{Arg}\det(M)$ but since the observable
combination $\theta+\rm{Arg}\det(M)$ remains constant we can use them to eliminate $\theta$.} $\sim \theta$,
\begin{equation}
\langle \bar{u}_{\rm{L}}u_{\rm{R}}\rangle=|\langle \bar{u}_{\rm{L}}u_{\rm{R}}\rangle|
\exp(\i(\alpha+c_{u}\theta)) \quad{\rm{and}}\quad
\langle \bar{d}_{\rm{L}}d_{\rm{R}}\rangle=|\langle \bar{d}_{\rm{L}}d_{\rm{R}}\rangle|
\exp(\i(-\alpha+c_{d}\theta)),
\end{equation}
where $c_{u}$ and $c_{d}$ are constants.
Taking the expectation value of the quark mass terms in the Lagrangian and approximating
$\langle \bar{u}u\rangle\approx \langle \bar{u}u\rangle$ one finds a contribution to the
effective potential,
\begin{equation}
V\sim |\langle \bar{u}u\rangle|\left[m_{u}\cos(\alpha+c_{u}\theta)+m_{d}\cos(\alpha-c_{d}\theta)\right],
\end{equation}
with the  up and down quark masses $m_{u}$ and $m_{d}$.
Expanding to second order in the fields and using properly normalized pion, $\pi_{0}$, and axion,  $a$, fields,
\begin{equation}
\alpha=\frac{\pi_{0}}{f_{\pi}},\quad \theta=\frac{a}{f_{a}}
\end{equation}
one finds
\begin{equation}
f^{2}_{\pi}m^{2}_{\pi}=(m_{u}+m_{d})\langle \bar{u}u\rangle
\end{equation}
and\footnote{We neglect a function $F(m_{u}/m_{d})$ which is of order unity.}
\begin{equation}
\label{massaxion}
m^{2}_{a}\sim \frac{f^{2}_{\pi}m^{2}_{\pi}}{f^{2}_{a}}.
\end{equation}
Here, $f_{\pi}$ is the pion decay constant and $f_{a}$ is the axion decay constant which characterize
the scale of (spontaneous) chiral and Peccei-Quinn symmetry breaking, respectively.

\begin{figure}
\begin{center}
\includegraphics[width=9cm]{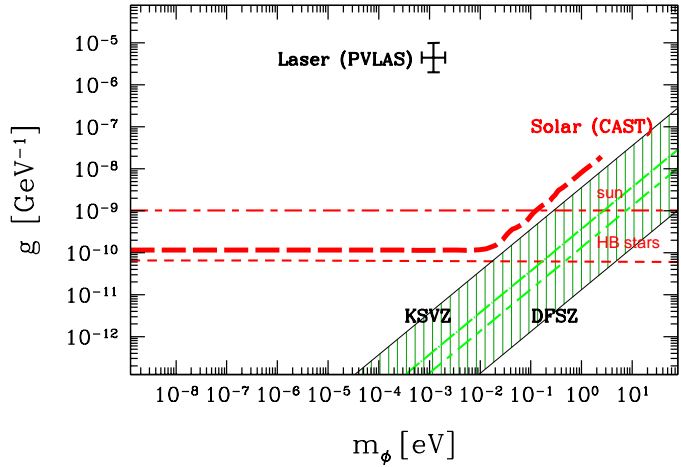}
\vspace{-0.7cm}
\end{center}
\caption{Mass-coupling plane for axion-like particles. The cross denotes the coupling required to explain PVLAS. The green
shaded area gives the typical range allowed for true axions and the green lines denote two prominent
examples of axion models \cite{Kim:1979if,Shifman:1979if,Dine:1981rt,Zhitni}.
The red lines represent the bounds from astrophysics
(everything above the lines is excluded). The bounds labelled ``sun'' and ``HB stars'' come from
energy loss considerations \cite{Frieman:1987ui,Raffelt:1985nk,Raffelt:1987yu,Raffelt:1996}, the ``CAST'' bound comes from the non-observation of axion-like particles in the
CAST experiment that aims to reconvert ALPs generated in the sun into photons \cite{Zioutas:2004hi}.}
\label{bounds}
\end{figure}

The, for our purposes particularly important, coupling to the electromagnetic field strength arises via
a similar triangle diagram as the coupling to the gluonic field strength.
Naturally it involves the elctromagnetic coupling $\alpha$,
\begin{equation}
\label{couplingaxion}
{\rm{coupling~to~two~photons}}\sim \alpha\frac{a}{f_{a}} F^{\mu\nu}\tilde{F}_{\mu\nu}.
\end{equation}
Please note that $f_{\pi}$ and $m_{\pi}$ are known and we have only one free parameter, $f_{a}$,
in Eqs. (\ref{massaxion}) and (\ref{couplingaxion}).
Mass and coupling of the axion are related.
Therefore, axions are confined to a line in the mass-coupling plane.
Of course our analysis neglected model dependent factors of order unity
and the line becomes a (narrow) band as shown in Fig. \ref{bounds}.

As we can see from Fig. \ref{bounds} the mass and coupling required to explain the PVLAS observation is
orders of magnitude outside the range allowed for axion models. Therefore, we cannot use the true axions invented
to solve the strong CP problem to explain PVLAS. In the following we will call light particles coupled to two photons
but without the mass-coupling relation axion-like particle or ALPs.

\section{Astrophysical Bounds}\label{astro}
A serious challenge for the particle interpretation of PVLAS are the extremely tight bounds on the couplings
of light particles to photons. In this section we will sketch how those bounds arise and in the
following Sect. \ref{evasion} we discuss some ideas from
the recent literature to circumvent them.
\subsection{Energy loss bound}
The production processes depicted in the upper half of Figs. \ref{alpconversion}
can also occur in the plasma in the central regions of the sun. Photons of the solar plasma interact with the electromagnetic fields
provided by the ions and electrons in the plasma and are converted into ALPs.
If the coupling of these particles is, however, sufficiently small they immediately leave the sun without
further scattering -- each one carrying away an energy of the order of $T_{\rm{center}}\sim 10^{7}\,\rm{K}\sim 1\,\rm{keV}$.
This is in stark contrast to photons. Photons typically scatter millions and millions of times
until they reach the surface of the sun. This has two effects. First it simply takes longer for a photon to leave
the sun, slowing down the energy transport via photons. Second on their long way to the surface the photons loose energy and
stay thermalized. They leave the sun only with energies of the order of the surface
temperature $T_{\rm{surface}}\sim 6000\, \rm{K}\sim 1\,\rm{eV}$ instead of having energies typical for the center of the sun
$\sim 1\,\rm{keV}$ resulting in a further slowdown of the energy transport.

Overall this makes the energy loss channel via light particles weakly coupled to photons extremely efficient when
compared to the standard one via photons. For example, for ALPs one finds \cite{Raffelt:1996}
\begin{equation}
\label{loss}
L_{\rm{ALP}}\sim
10^{7}\left(g\,10^{5}\rm{GeV}\right)^{2}L_{\gamma},
\end{equation}
where $L_{\rm{ALP}}$ is the energy loss via ALPs and $L_{\gamma}$ the one via photons.

To compensate for the additional energy loss channels the sun would have to burn fuel at a much
faster rate\footnote{If we could switch on such an energy loss channel now, the center of the sun would cool a
bit. Consequently it would contract until the rate of nuclear fusion has risen enough to compensate for
the additional energy loss.}.
Without additional light particles the sun has enough nuclear fuel for about $10^{10}\,\rm{years}$.
Using Eq. (\ref{loss}) we can easily estimate that in the presence of ALPs with a coupling strong
enough to explain PVLAS the sun would have fuel for at most a few 1000 years.
Clearly in conflict with observation.
A more detailed analysis strengthens the bound to the limit \cite{Frieman:1987ui,Raffelt:1996} depicted in Fig. \ref{bounds}.
Including data from so called horizontal branch (HB) stars one can improve it even more
\cite{Raffelt:1985nk,Raffelt:1987yu} (cf. Fig. \ref{bounds}).

For millicharged particles the production process in the upper half of Fig. \ref{milliconversion}
can be used for a similar argument leading to \cite{Davidson:1991si,Mohapatra:1990vq,Mohapatra:1991as,Davidson:1993sj,Davidson:2000hf},
\begin{equation}
\epsilon\lesssim 10^{-14}\quad\rm{for}\quad m_{\epsilon}\lesssim 1\,\rm{keV}.
\end{equation}
\subsection{CAST Bound}
For ALPs one can also try to reconvert the ALPs generated in the sun \cite{Sikivie:1983ip}. Basically this
uses the inverse of the production process of Fig. \ref{alpconversion}. The CAST collaboration \cite{Zioutas:2004hi}
has set up such an experiment using an LHC test magnet to provide the background magnetic field.
Looking in the direction of the sun one would expect to regenerate photons with energies of a few
$\rm{keV}$ (temperature of the sun).
So far CAST has found no signal above the background. This gives the ``CAST'' bound in Fig. \ref{bounds}.

For millicharged particles the CAST idea does not work because it is highly unlikely that two millicharged
particles would meet again inside the CAST magnet.
\section{Evading Astrophysical Bounds}\label{evasion}
Inspired by the PVLAS result and the apparent conflict with astrophysical bounds several models to evade the astrophysical
bounds have been proposed. We will start by discussing a few ideas that apply to ALPs then we move briefly
to millicharged particles until we come to a few entirely different ideas to explain PVLAS without violating
astrophysical bounds.

\subsection{Trapping}
One central property that leads to the tightness of astrophysical bounds is the fact that ALPs, once produced, immediately
leave the sun. So one might want to endow the ALPs with some additional interaction that prevents their escape
from the sun basically trapping them in the sun (at least for some time).
This would dramatically reduce the energy loss. Moreover, it would also prevent detection via CAST because the ALPs
would loose energy in scattering processes making them undetectable by CAST which is sensitive only
to ALPs with energies $\gtrsim 1\,\rm{keV}$.
The basic concept for this mechanism was already sketched in \cite{Masso:2005ym} and models were presented in~\cite{Jain:2005nh,Jain:2006ki}.

This idea has one essential difficulty. To achieve a sufficient amount of scattering of ALPs within the sun
one needs a fairly large interaction making it difficult to avoid detection in other experiments and observations.
However, so far no such conflicts have been found for the models~\cite{Jain:2005nh,Jain:2006ki} and they remain viable.

\subsection{Suppressed Production}
An alternative to trapping is to find a way to suppress the production of ALP's inside the sun.
Models of this type are based on the fact that the environment inside the sun is different from
the lab environment of PVLAS.
The various proposed models make use of different parameters that are higher in the sun than in the lab:
\begin{itemize}
\item{} The typical virtuality of the process producing the ALPs \cite{Masso:2005ym,Masso:2006gc}
\item{} The plasma frequency in the solar plasma \cite{Masso:2005ym,Masso:2006gc}
\item{} Temperature \cite{Mohapatra:2006pv}
\item{} Density \cite{Jaeckel:2006id}.
\end{itemize}

The problem with these models is that they typically involve at least one unnaturally small parameter.
This difficulty is enhanced by the fact that it is not enough to suppress the production of ALPs in the center of the
sun but one has to achieve suppression also in the outer parts of the sun where temperature, density etc. are
significantly smaller than in the center \cite{Jaeckel:2006id}\footnote{This can actually be turned into an advantage. Some
of the parameters are so small that we may be able to recreate them in the laboratory thereby allowing for
a direct test of those models in the lab.}.
\subsection{Millicharged Particles}
As we already discussed pair production  of millicharged particles can provide an alternative \cite{Gies:2006ca}
to the single production
of neutral particles coupled to two photons.
However, we already saw in Sect. \ref{astro} that they face similar problems from astrophysical bounds.
The most straightforward way to avoid this problem is, again, to switch off production. Indeed the
model \cite{Masso:2006gc} already involves millicharged fermions and is actually simplified by removing the additional axion-like particle.

A nice feature of this model is that millicharged particles occur naturally
in extensions of the standard model \cite{Holdom:1985ag,Abel:2003ue,Batell:2005wa}.
For example, in D-brane models of string theory one can find the right representations and
a sufficient amount of kinetic mixing to explain PVLAS \cite{Abel:2006qt}.
Nevertheless, there remains an unexplained small mass scale of the order of $\rm{meV}$ in these models, too.
\subsection{Other proposals}
In addition to the above there have been a few proposals that involve neither spin-0 bosons coupled to two
photons nor millicharged particles.

One possibility is spacetime noncommutativity \cite{Chaichian:2005gh}.
This idea avoids additional light particles
and therefore also their problems with astrophysical bounds. However, a recent careful
study \cite{Chatillon:2006rn} suggests that
spacetime noncommutativity produces only an ellipticity and cannot explain the rotation signal of PVLAS.

Another possibility tried to employ an additional vector particle \cite{Antoniadis:2006wp}.
This also produces an ellipticity but, unfortunately, probably fails to produce the required amount of rotation.

Finally, there has been a claim \cite{Mendonca:2006pg} that the PVLAS result can be explained within
standard QED if the (slow) rotation
of the background magnetic field is taken into account. A careful wave-propagation study \cite{Adler:2006zs,Biswas:2006cr} has, however, shown that this is ruled out.
\section{Experimental Tests}
Confirmation of the PVLAS result and distinguishing between the different proposed models
requires further experiments.
One way is to make an independent experiment testing and improving the measurements of PVLAS.
One experiment currently under way is the Q\&A experiment which recently has published first data
\cite{Chen:2006cd} but
more data from different sources is expected in the near future \cite{Pugnat:2005nk,Rizzo:Patras,Heinzl:2006xc}.
A detailed analysis of the data from such optical experiments may even allow to distinguish between axion-like
and millicharged particles~\cite{Ahlers:2006iz}.

However, an experiment of the PVLAS type is a disappearance experiment (the produced particles are not detected).
It would be nice to really have some way to actually detect the proposed particles.
In this section we want to sketch two experiments that are sensitive
to axion-like particles and millicharged particles, respectively.

\subsection{Light Shining through Walls}
To test for the existence of spin-0 bosons coupled to two photons one can do a so
called ``light shining through a wall experiment''
\cite{Anselm:1986gz,Gasperini:1987da,VanBibber:1987rq,Ruoso:1992nx} as sketched in Fig. \ref{fig:ph_reg}.
Here, the photons are first converted into ALPs using a strong magnetic field.
Then the photon beam is stopped by a thick wall.
The weakly interacting ALPs, however, pass through this wall without any problems. Behind the wall
another magnetic field is used to reconvert the ALPs into photons via the inverse production process.
The photons (coming ``out of the wall'') are then detected.
This is a very clean detection experiment. The photons coming out of the wall have the same frequency as the laser making it easier to distinguish them from the background.
Moreover the number of photons coming out of the
wall is directly correlated to the strength of the laser, e.g. for a pulsed laser one can then use the resulting time
dependence of the signal to further improve the signal/noise ratio.

\begin{figure}
\begin{center}
\includegraphics[width=8.5cm]{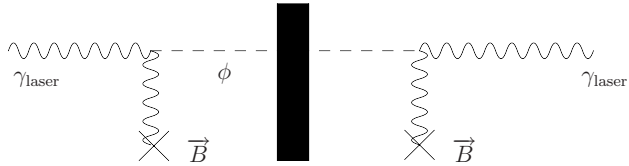}
\end{center}
\vspace{-0.5cm}
\caption{Schematic view of a ``light shining through a wall'' experiment.
(Pseudo-)scalar production through photon conversion in a magnetic field (left), subsequent travel
through a wall, and final detection through photon regeneration
(right). }
\label{fig:ph_reg}
\end{figure}

Currently, at least four such experiments are in advanced planning/testing stages
(for a recent review see \cite{Ringwald:2006rf}).
One at DESY \cite{ALPS}, one at Jefferson lab \cite{Baker:Patras},
one by the BMV collaboration \cite{Rizzo:Patras} and one at CERN \cite{Pugnat:2005nk}.
The PVLAS collaboration themselves also plan an extension of their experiment \cite{Cantatore:Patras}.

\subsection{Dark Currents Shining through a Wall}
Unfortunately, a light shining through a wall experiments will not work for millicharged particles.
Once produced the particle and the antiparticle will move away from each other since they experience
opposite forces in the magnetic field due to their opposite charges\footnote{In addition the produced
particle and antiparticle have typically opposite momenta along the magnetic field lines separating them
also in this direction.}.
They will simply not meet again behind the wall and reconversion does not occur (most of them will not even make it to the wall because the magnetic field forces them on circular trajectories).

A way out is to use another QED process: Schwinger pair production \cite{Schwinger:1951nm}.
In a strong electric field (without an additional laser) vacuum
pairs of charged particles that are created at a sufficiently large distance,
\begin{equation}
 d>d_{\rm{pair}}=\frac{2m_{\epsilon}}{\epsilon e \mathbf{E}}
\end{equation}
have gained more energy from the electric field than it cost to produce them and
it is energetically favorable to produce such pairs. Roughly speaking this is a tunneling process
where a barrier of height $2m_{\epsilon}$ (the energy to produce the pair) and
width $d_{\rm{pair}}$ (the distance needed to gain enough energy in the electric field)
must be overcome.
The pairproduction rate is high if the electric field is large enough such that $d_{\rm{pair}}/2$ is smaller than the Compton wavelength of the particles $\lambda_{C}=1/m_{\epsilon}$,
\begin{equation}
\mathbf{E}>\mathbf{E}_{\rm{crit}}=\frac{m^2_{\epsilon}}{\epsilon e}.
\end{equation}

\begin{figure}
\begin{center}
\vspace{1cm}
\includegraphics[angle=-90,width=8.5cm]{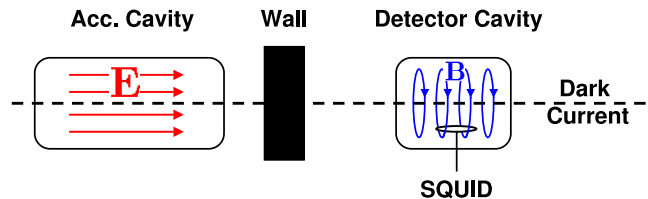}
\end{center}
\vspace{-0.3cm}
\caption{
Schematic set up for a ``dark current shining through a wall''
experiment \cite{Gies:2006hv}. The alternating dark current (frequency $\nu$), comprised of the produced millicharged particles
(dashed line), escapes from the accelerator cavity  and traverses also a
thick shielding (``wall''), in which the conventional dark current of electrons
is stopped. The dark current induces a magnetic field in a resonant (frequency $\nu$) detector cavity
behind the wall, which is detected by a SQUID.}
\label{acdcdetector}
\end{figure}

Since the produced particles typically have momenta along the lines of the electric field for the positive charge particle and in the opposite direction for the negative counterpart a current is generated.
This current can be measured with a setup shown in Fig. \ref{acdcdetector}. Using existing
accelerator cavities as source of the electric field such an experiment could be build in the near future
allowing to test the millicharged particle interpretation of PVLAS \cite{Gies:2006hv}.

\section{Conclusions}

Motivated by the surprising observation of a non-vanishing rotation of linear polarized laser light in
a magnetic field by the PVLAS collaboration we have reviewed light particles coupled to photons.

While standard QCD axions are unable to explain PVLAS similar light particles coupled to two photons, axion-like particles, may do so. Another candidate are light particles with a small electric charge.
If confirmed the PVLAS signal would be the first direct evidence for physics beyond the standard model.

Both particle interpretations of PVLAS face serious problems from astrophysical bounds
(so far other explanation attempts have not been succesful).
Models to avoid the astrophysical bounds can be constructed but typically require some fine-tuning.

Several experiments to test PVLAS and its particle interpretations are currently under way or will/can be build in the near future.

The existence of light particles coupled to photons would open a window to
physics beyond the standard model via low energy experiments with photons that is complementary to conventional accelerator experiments.
\section*{Acknowledgements}
The author wishes to thank the organizers of the symposium on ``Large TPCs for low energy rare
event detection'' in Paris. Many thanks go also to S.~Abel, M.~Ahlers, H.~Gies, V.V.~Khoze, E.~Masso, J.~Redondo,
A.~Ringwald and F.~Takahashi for interesting discussions and productive collaboration.


\end{document}